\documentclass[aps,prl,twocolumn,groupedaddress]{revtex4}

\usepackage{graphicx} 
\usepackage{dcolumn}
\usepackage{bm}

\begin{document}

\title{Very Unusual Magnetic Properties in Multi-walled Carbon 
Nanotube Mats} 
\author{Guo-meng Zhao$^{*}$ and Pieder Beeli} 
\affiliation{Department of Physics and Astronomy, California State 
University, Los Angeles, CA 90032, USA}


\begin{abstract}

We report magnetic measurements up to 1100 K on a multi-walled  
carbon  nanotube mat sample using a Quantum Design vibrating sample 
magnetometer. In an ultra-low field  ($H$ = $-$0.02 Oe), we find a 
very 
large paramagnetic susceptibility (up to 12.7$\%$ of 1/4$\pi$) at
1100 K and a
very large diamagnetic susceptibility (at least 8.4$\%$ of 
$-$1/4$\pi$) at 482 K. 
A small magnetic field (2.1 Oe) completely suppresses the diamagnetic 
susceptibility at 482 K and reduces the paramagnetic susceptibility 
at 1100 K by a factor of over 20. We rule out explanations based on 
magnetic contaminants, instrument artifacts, and  orbital diamagnetism. The magnetic data  are inconsistent with any known 
physical phenomena except for granular 
superconductivity. The present results suggest the 
existence of an unknown new physical phenomenon or 
superconductivity with an ultra-high transition temperature.

\end{abstract}

\maketitle

The low-energy physics of an individual metallic single-walled carbon 
nanotube is shown to be equivalent to a two-leg Hubbard ladder 
\cite{Balent}, which can be described by a Luttinger liquid with an 
effective interaction parameter $K_{\rho+}$ (Refs.~\cite{Kiv,Orig}).  
Single-walled nanotube (SWNT) bundles and individual multi-walled 
nanotubes (MWNTs) are then equivalent to quasi-one-dimensional 
(quasi-1D) arrays of two-leg Hubbard ladders, which should be ideal 
candidates for high-temperature superconductivity \cite{Kiv,Orig} 
especially if $K_{\rho+}$ $\geq$ 1.5 (Ref.~\cite{Orig}). The $K_{\rho+}$ value of a two-leg ladder can 
be determined from the temperature dependence of the resistance $R$, 
which follows $R$ $\propto$ $T^{2K_{\rho+}-2}$ (Ref.~\cite{Orig}).  The 
on-tube resistance is found \cite{Zhaorev3} to follow $R_{tube}$ 
$\propto$ $T^{1.72}$ for a SWNT with a diameter $d$ = 1.5 nm 
(corresponding to $K_{\rho+}$ = 1.86) and $R_{tube}$ $\propto$ $T^{1.48}$ 
for another SWNT with $d$ = 1.7 nm (corresponding to $K_{\rho+}$ = 
1.74).  The large $K_{\rho+}$ values imply that the 
long-range repulsive Coulomb interaction in these samples is effectively screened out
so that strong electron-phonon interactions with the radial 
breathing mode \cite{Egg03} and other phonon modes may 
lead to a large effective attractive interaction.  Therefore, SWNT 
bundles and MWNTs are ideal candidates for high-temperature superconductivity.

Very recently, quasi-1D superconductivity at about 12 K has been reported 
\cite{Take} in a short MWNT array where the normal-state resistance $R_{N}$ 
$\propto$ $T^{-0.8}$.  Such a temperature dependence of $R_{N}$ 
might suggest that $K_{\rho+}$ $\leq$ 0.6, leading to  $d$-wave 
superconductivity \cite{Orig}.  Since any small disorder can suppress superconductivity 
dramatically in the case of $K_{\rho+}$ $<$ 1 (Ref.~\cite{Orig}), the 
intrinsic superconducting transition temperature in disorder-free MWNT arrays should 
be much higher.  On the other hand, the resistance data \cite{Ebb} of an individual MWNT can 
be quantitatively explained in terms of quasi-1D superconductivity 
with a mean-field transition temperature $T_{c0}$ $\simeq$ 260 K 
\cite{Zhaorev3,Zhaorev4}.  The normal-state on-tube resistance of this 
individual MWNT appears to follow a linear-$T$ dependence 
\cite{Zhaorev3,Ebb,Zhaorev4}, suggesting that $K_{\rho+}$ $\simeq$ 1.5.  
The larger $K_{\rho+}$ ($>$ 1) and $T_{c0}$ values found for this MWNT indicate that the 
long-range repulsive Coulomb interaction is more effectively screened 
out and/or the doping is more optimal.

Here we report magnetic measurements up to 1100 K on a MWNT mat sample 
using a Quantum Design vibrating sample magnetometer (VSM).  In an 
ultra-low field ($H$ = $-$0.02 Oe), we find a very large paramagnetic 
susceptibility (up to 12.7$\%$ of 1/4$\pi$) at
1100~K 
and a very large diamagnetic susceptibility (at least 8.4$\%$ of 
$-$1/4$\pi$) at 482~K.  A small magnetic field (2.1 Oe) completely 
suppresses the diamagnetic susceptibility at 488 K and reduces the 
paramagnetic susceptibility at 1100~K by a factor of over 20.  The 
results are inconsistent with any known physical phenomena except for granular 
superconductivity with $T_{c0}$ $>$ 1100~K.

A purified MWNT mat sample is obtained from SES Research of Houston.  The 
sample was prepared by chemical vapor deposition (CVD) using an iron 
catalyst.  By burning off carbon related materials in air (thermal 
gravimetric analysis), we find the weight of the residual to be 1.0$\%$.  We determine the relative
metal 
contents of the residual using a Perkin-Elmer Elan-DRCe inductively 
coupled plasma mass spectrometer (ICP-MS). From the ICP-MS result and 
the percentage of the residual, we obtain the metal-based impurity 
concentrations (ppm in weight): Na =1210.3, Mg = 80.156, Al = 1645.5, 
Si = 184.43, K = 233.51, Ca = 1129.9, Sc = 2.027, Ti = 5.4618, V = 
1.7611, Cr = 33.634, Mn = 10.953, Fe = 5225.1, Co = 2.7628, Ni = 
10.354, and Cu = 224.24. The main magnetic impurities are 
Fe$_{2}$O$_{3}$, as 
indicated by the energy dispersive x-ray (EDX) analysis. The Co concentration in this sample is less than 
3 ppm, which is negligibly small.

Extensive magnetic measurements by both us and the staff of Quantum 
Design (up to 1100 K) on a control sample of Er$_{2}$O$_{3}$ show 
that$-$after 
correcting for an offset moment of (4.4-5.6)$\times$10$^{-7}$ emu 
due to the 
interaction between the VSM drive head and the pick-up coils$-$the 
absolute accuracy of the measurement is better than 1.0$\times$10$^{-7}$ 
emu at low fields (0.06-4.21 Oe) \cite{Zhao}.  The data shown below
are corrected for temperature-dependent offset moment: (5.5$\times$10$^{-7}$ $-$ 
4.4$\times$10$^{-11}T$) emu, which is determined from magnetic 
measurements on the control sample of  Er$_{2}$O$_{3}$ cooled from $T$ = 
1100 K to 300 K in a field of $-$0.018 Oe. The data are smoothed using cubic spline interpolation 
with a smoothing parameter $\Lambda$ = 4. The stable low fields  are obtained after the ultra 
low-field process leaves a slightly negative field at the 
sample position.  Then the field is biased using a secondary coil wound on 
the vacuum sleeve and driven with a dc current.  The field profile is 
obtained with the high-resolution flux gate whose absolute magnitude 
is verified with the paramagnetic standard sample.

Figure 1 shows the temperature dependence of the initial warm-up and 
cool-down 
susceptibilities in a field of $-$0.02 Oe for a 12.3 mg MWNT mat 
sample.  The virgin sample was inserted into the sample chamber with the field of 
$-$0.02 Oe without going through the linear motor used for vibrating 
the sample.  The maximum field the sample has experienced is the 
earth-field (about $-$0.5 Oe) before it experiences the field of 
$-$0.02 Oe.  Therefore, any remanent 
magnetization due to 
superconductivity and/or magnetic impurities should be negative when 
the field is reduced from $-$0.5 Oe to $-$0.02 Oe. This will lead to an underestimate of the magnitude of the diamagnetic 
susceptibility because of the negative field ($-$0.02 Oe).

\begin{figure}[htb]
\includegraphics[height=7cm]{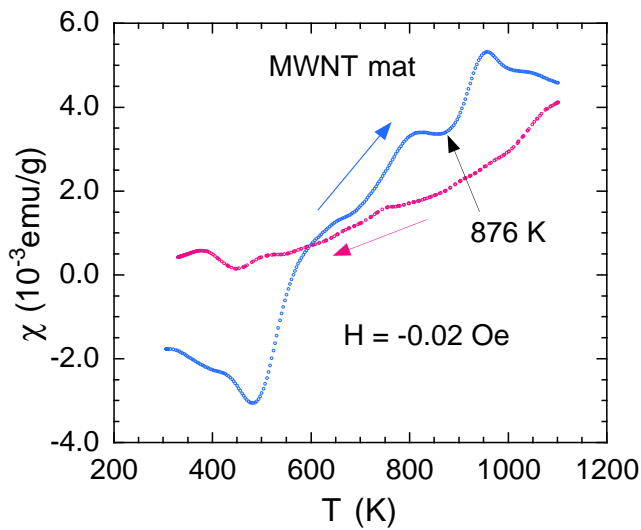} 
\caption[~]{The temperature 
dependence of the initial warm-up and cool-down susceptibilities in a 
field of $-$0.02 Oe for a 12.3 mg MWNT mat sample.  The initial 
warm-up susceptibility at 482 K has a large negative value (about 
$-$3.05$\times$10$^{-3}$ emu/g), corresponding to about 8.4$\%$ of
the full Meissner
effect ($-$1/4$\pi$). }
\end{figure}

From the warm-up data, we clearly see a negative peak at about 482 K 
where the diamagnetic susceptibility is $-$3.05$\times$10$^{-3}$ 
emu/g, corresponding to about 8.4$\%$ of $-$1/4$\pi$ given that the specific weight of 
MWNTs is about 2.17 g/cm$^{3}$. Such a strong diamagnetism 
cannot be explained by the orbital diamagnetism of MWNTs. This is because the magnitude 
of the total diamagnetic susceptibility at room 
temperature  is 2.6$\times$10$^{-6}$ emu/g in mechanically 
ground MWNTs \cite{Tsui,Jin}, which should be the upper 
limit of the magnitude of the orbital diamagnetic susceptibility.  It is clear 
that the orbital diamagnetic susceptibility of MWNTs is over {\em 
three orders of magnitude} smaller than the 
observed diamagnetic susceptibility at 482 K.  Moreover, the orbital 
diamagnetic susceptibility is predicted to be field independent 
\cite{Lu}, in contrast to a very strong field dependence of the 
diamagnetic susceptibility (see below).

It is known that a negative susceptibility peak can 
occur in granular superconductors for the warm-up measurement due to 
flux relaxation \cite{Hyun}.  If the MWNT mat sample is a granular 
superconductor with an intergrain Josephson coupling temperature 
($T_{cJ}$) above room temperature, some flux lines can be trapped in 
the regions of Josephson weak links at 300 K when the field is reduced 
from $-$0.5 Oe to $-$0.02 Oe.  As the sample is warmed, the trapped 
flux lines will relax and move out of the sample, leading to a 
negative susceptibility peak slightly below $T_{cJ}$ where the flux 
pinning force is greatly reduced due to the weakening of Josephson 
coupling \cite{Hyun}.  If this interpretation is relevant, $T_{cJ}$ of 
this MWNT mat sample is at least 482~K.

It is also interesting that there is a pronounced shoulder feature 
between 800 K and 900 K, which should arise from the ferrimagnetic 
ordering of $\gamma$-Fe$_{2}$O$_{3}$ impurities whose Curie 
temperatures 
are between 863-950 K (Ref.~\cite{Dun}).  If we subtract the ferrimagnetic 
component of the susceptibility, the magnitude 
of the intrinsic diamagnetic susceptibility at 482 K will be much larger 
than 3.05$\times$10$^{-3}$ emu/g.  Moreover, this cool-down 
susceptibility in $-$0.02 Oe for this CVD prepared sample is nearly 
the same as that in $+$0.03 Oe for an 
arc-discharge prepared sample \cite{Zhao} although the two samples have very different masses (12.3 mg 
versus 41 mg) and magnetic contaminations (5500 ppm versus 100 ppm), and 
are respectively subject to negative 
and positive fields ($-$0.02 Oe versus $+$0.03 Oe).  Such consistency 
rules out explanations based on instrument artifacts and 
magnetic contaminations.

Figure~2a shows susceptibility versus temperature for our MWNT mat in a field of 
$-$0.02 
Oe through thermal cycling.  It is evident that the susceptibility 
becomes more diamagnetic upon thermal cycling.  This is consistent 
with the fact that the trapped flux lines are progressively released 
upon heating up to 1100 K where the flux pinning force is much 
weaker. Similar behavior was observed \cite{Hyun} in polycrystalline 
Nb$_{3}$Sn, as shown in Fig.~2b. 

Figure 3 shows both zero-field (ZFC) and field-cooled (FC) 
susceptibilities in a field of 2.1 Oe. It is apparent that the ZFC 
and FC susceptibilities show a significant difference throughout the 
whole temperature region. Moreover, when compared with the 
susceptibility in the field of $-$0.02 Oe, the diamagnetism in the 
field of 2.1 Oe disappears and the paramagnetic susceptibility at 
1100 K is reduced by a factor of over 20. Although the temperature 
dependence of the FC susceptibility in such a low field has not been 
seen in any known ferromagnets or ferrimagnets, the result in Fig.~3 
alone would also be consistent with a magnetic transition above 1100~K.

\begin{figure}
\includegraphics[height=7cm]{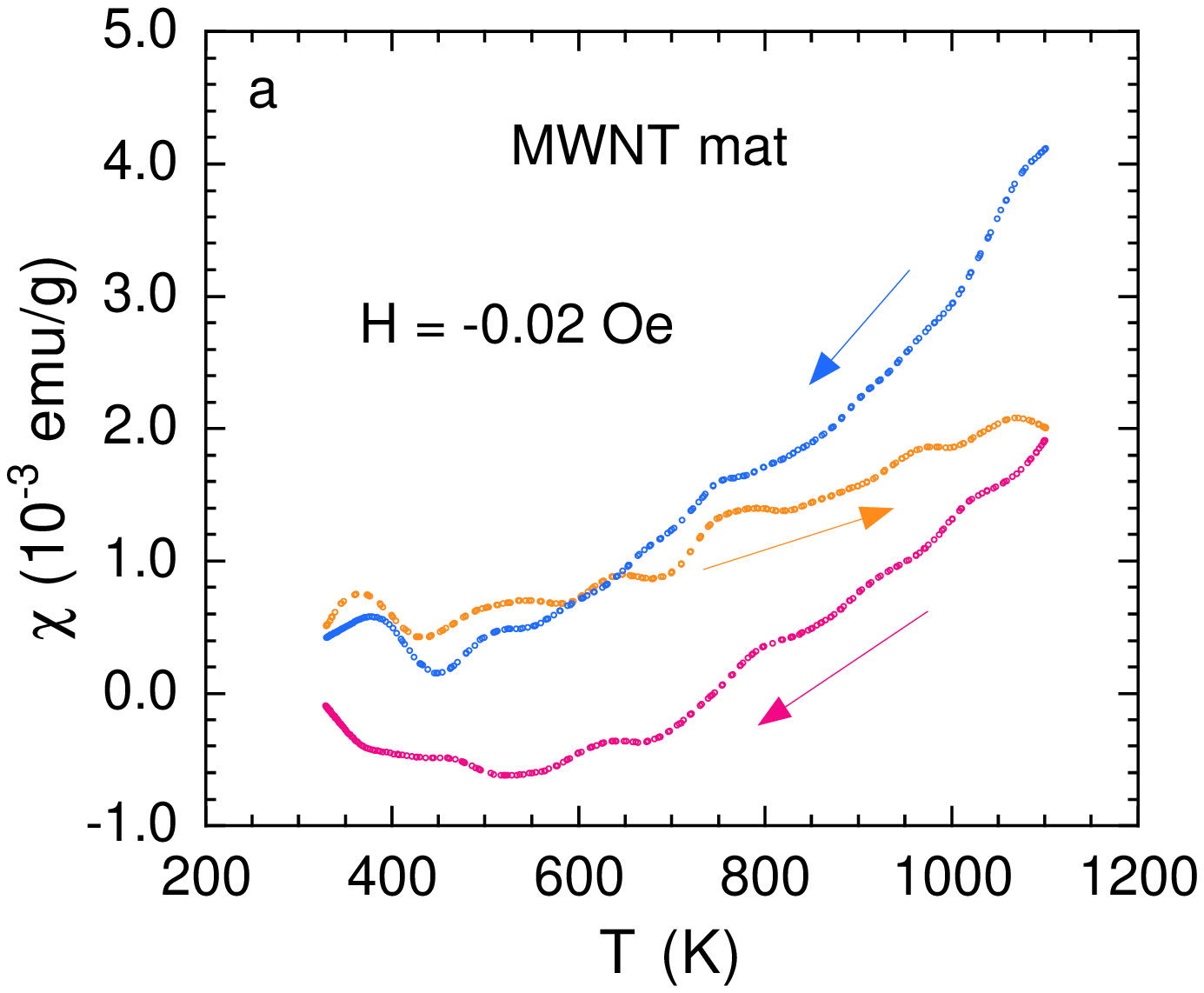} 
\includegraphics[height=7cm]{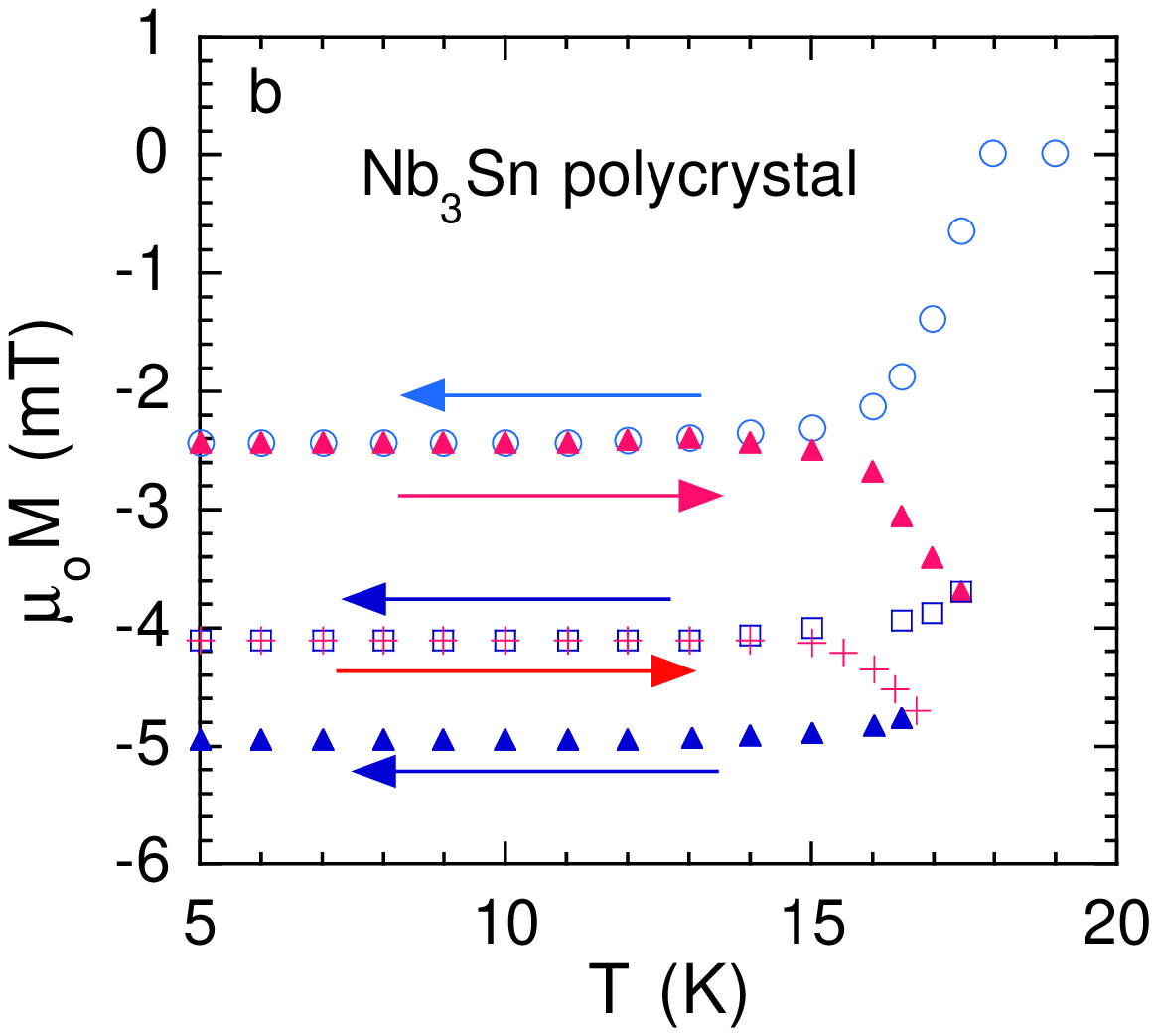} 
\caption[~]{ 
(a) The susceptibility versus temperature in a field of $-$0.02 Oe 
through thermal cycling for the MWNT mat sample.  (b) The 
magnetization versus temperature in a field of 100 Oe through thermal 
cycling for a polycrystalline Nb$_{3}$Sn.  The data are reproduced 
from \cite{Hyun}.}
\end{figure}

The very large paramagnetic susceptibility (about 
4.6$\times$10$^{-3}$ emu/g) at 
1100 K for the MWNT mat sample in the ultra-low field (see Fig.~1) cannot arise 
from magnetic contaminants.  If this were caused by magnetic 
contaminants with the Curie temperatures higher than 1100 K, the 
candidates would only be Co and its alloys. For materials 
contaminated with magnetic impurities having very large intrinsic 
susceptibilities, the initial low-field susceptibility is independent 
of the intrinsic susceptibility and depends only on the shape 
(demagnetization factor) and concentration of the contaminant 
\cite{Jack}.  
The shape effect is averaged out through a random orientation so that 
the average demagnetization factor $N$ should be about 1/3 and the 
low-field susceptibility should be 1/(4$\pi N$) = 0.24 emu/cm$^{3}$.  
Given that the 
Co specific weight is 8.9 g/cm$^{3}$, the low-field susceptibility of 
pure 
Co particles is calculated to be 0.027 emu/g. The precise elemental 
analysis on the non-carbon residual of the MWNT sample indicates that 
the Co concentration is below 3 ppm. The initial low-field 
susceptibility for this ultra-low Co concentration is therefore 
estimated to be 3$\times$10$^{-6}$$\times$0.027 emu/g = 
8.1$\times$10$^{-8}$ emu/g, which is about 
{\em five orders of magnitude} smaller than the measured value.
  
\begin{figure}[htb]
\includegraphics[height=7cm]{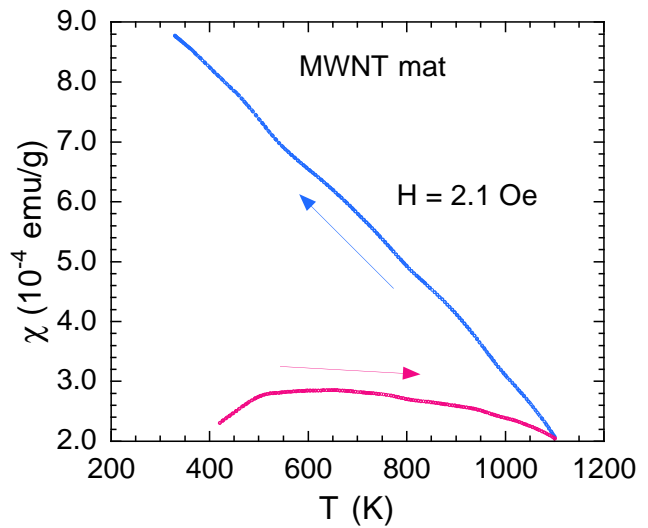} 
\caption[~]{The 
zero-field (ZFC) and field-cooled (FC) susceptibilities in a field of 
2.1 Oe for the MWNT mat sample.}
\end{figure}

However, one might argue that the very large paramagnetic 
susceptibility at 1100 K in the ultra-low field might originate from 
a ferromagnetic/ferrimagnetic transition above 1100 K in an unknown 
carbon phase. Such a high Curie temperature in a carbon-related 
material, which does not have $d$ electrons, would be very remarkable. Magnetic 
properties of amorphous-like carbon are found to depend on the 
hydrogen concentration \cite{Mur}. For a hydrogen-rich phase, the 
Curie 
temperature is about 500 K and the saturation magnetization is about 10 
emu/g (Ref.~\cite{Mur}).  Polymerized fullerenes \cite{Mak} and 
proton-irradiated graphite \cite{Esq} also show 
ferromagnetic/ferrimagnetic transitions above room temperature.  
Thus, it is not impossible that the very large paramagnetic 
susceptibility at 1100 K arises from a ferromagnetic/ferrimagnetic 
ordering above 1100 K in an unknown carbon phase.  However, the very 
rapid drop in the paramagnetic susceptibility with increasing field 
(e.g., 4.6$\times$10$^{-3}$ emu/g in $-$0.02 Oe and 2$\times$10$^{-4}$ 
emu/g in 2.1 Oe) in such a low field region ($\leq$2.1 Oe) is difficult 
to explain in terms of {\em any} known ferromagnetic/ferrimagnetic transition.  
For normal ferromagnets/ferrimagnets, the initial low-field 
susceptibility {\em increases} with increasing field in the field range less 
than 2.5 Oe \cite{Jack,Hro}.  Even in the weak ferromagnet 
RuSr$_{2}$GdCu$_{2}$O$_{8}$, the susceptibility in the whole 
temperature region below the Curie temperature {\em increases}  with 
increasing field for $H < $ 2.5 Oe \cite{Bern}.  It is even more 
difficult to explain the large diamagnetic susceptibility and the 
negative susceptibility peak feature at about 482 K in terms of 
ferromagnetism/ferrimagnetism.  Therefore, it appears unlikely that an 
unknown carbon phase with a ferromagnetic/ferrimagnetic transition 
above 1100 K is responsible for such unusual magnetic properties 
unless the ferromagnetic/ferrimagnetic phase would be very different 
from the conventional one.

Alternatively, the very large paramagnetic susceptibility at 1100 K 
and very large diamagnetic susceptibility at 482 K are consistent with 
the magnetic response in a granular superconductor \cite{Bar} with 
$T_{c0}$ $>$ 1100 K.  According to a theoretical simulation 
\cite{Bar}, the paramagnetic peak position occurs at about 
0.9$T_{c0}$, and the magnitude of the peak susceptibility decreases 
rapidly with increasing field in the ultra-low field region.  Our data 
are consistent with this theoretical prediction, but inconsistent with any 
known ferrimagnetic/ferromagnetic transition.  From Fig.~1, we also see that 
below the peak temperature, the susceptibility decreases with 
decreasing temperature and becomes negative at lower temperatures.  
This is also in agreement with the theoretical simulation for granular 
superconductors \cite{Bar}.  The large diamagnetic susceptibility at 
lower temperatures indicates that the diamagnetic Meissner effect 
outweighs the paramagnetic Meissner effect.  The absence of 
diamagnetism in the field of 2.1 Oe is consistent with the fact that 
2.1 Oe is much larger than the average intergrain lower critical field 
$H_{c1}$, which might be much lower than 0.02 Oe at 300 K.  In this case, the diamagnetic Meissner effect is 
overwhelmed by the paramagnetic Meissner effect.  In the high field 
range ($>$20 kOe), intergrain diamagnetic and paramagnetic Meissner 
effects will almost vanish so that the intragrain diamagnetic response 
dominates at low temperatures.  The intragrain diamagnetic 
susceptibility is small because the magnetic penetration depth is much 
larger than the grain sizes \cite{Zhaorev3,Zhaorev4}.

In summary, we have observed very unusual magnetic properties in MWNT 
mat samples.  The data are inconsistent with any known physical phenomena 
except for granular superconductivity with an ultra-high intragrain 
transition temperature ($>$ 1100 K).  The much higher $T_{c0}$ in the 
mat sample than that ($\sim$260 K) in the individual MWNT 
\cite{Ebb,Zhaorev3,Zhaorev4} may be due to more effective screening 
of the long-range Coulomb interaction, much less disorder, and more 
optimal doping in the mat samples. Alternatively, the very unusual magnetic 
properties observed in the MWNT 
mat samples might suggest the 
existence of an unknown new physical phenomenon, which has similar 
magnetic properties as granular superconductors.

\begin{acknowledgments}
We thank J.  O'Brien from Quantum Design 
for technical assistance.  We also thank G.  Gao and F.  M.  Zhou in 
the 
Department of Chemistry and Biochemistry at CSULA for the elemental 
analyses using ICP-MS.  We are grateful to Ivan Alvarado-Rodriguez at 
UCLA for the SEM analyses.  This research is partly supported by a 
Cottrell Science Award from Research Corporation.
\end{acknowledgments}

\end{document}